\documentclass[11pt,sort&compress]{elsarticle}

\journal{}
\usepackage{lipsum}
\usepackage{pgfplots}
\usepackage{amsfonts}
\usepackage{graphicx}
\usepackage{epstopdf}
\usepackage{tikz}
\usepackage{caption}
\usepackage{subcaption}
\usepackage{listings}
\usetikzlibrary{shapes.geometric, arrows}
\usepackage{epsf}
\usepackage{amsmath}
\usepackage{amssymb}
\usepackage{setspace}
\usepackage{enumerate}
\usepackage{float}
\usepackage{footnote}
\usepackage{scalefnt}
\usepackage{xfrac}
\usepackage{transparent}
\usepackage{rotate}
\usepackage{array}
\usepackage{tabu}
\usepackage[mediumspace,mediumqspace,squaren]{SIunits}
\usepackage{multirow}
\usepackage{enumitem}
\usepackage{tikz}
\usepackage{bm}
\usepackage{microtype}
\usepackage[margin=1in]{geometry}
\usetikzlibrary{patterns.meta}
\usepackage{outlines}
\usepackage{mathtools}
\usepackage{xspace}
\usepackage{tikz}

\usepackage{bm}

\usepackage{hyperref}
\usepackage[]{xcolor}

\hypersetup{
  colorlinks=true,
}

\definecolor{lightblue}{rgb}{0.63, 0.74, 0.78}
\definecolor{seagreen}{rgb}{0.18, 0.42, 0.41}
\definecolor{orange}{rgb}{0.85, 0.55, 0.13}
\definecolor{silver}{rgb}{0.69, 0.67, 0.66}
\definecolor{rust}{rgb}{0.72, 0.26, 0.06}

\colorlet{lightsilver}{silver!30!white}
\colorlet{darkorange}{orange!75!black}
\colorlet{darksilver}{silver!65!black}
\colorlet{darklightblue}{lightblue!65!black}
\colorlet{darkrust}{rust!85!black}

\usepackage[labelfont=bf]{caption}
\usepackage{cleveref}

\definecolor{dkgreen}{rgb}{0,0.6,0}
\definecolor{gray}{rgb}{0.5,0.5,0.5}
\definecolor{mauve}{rgb}{0.58,0,0.82}

\definecolor{RYB1}{rgb}{0.63, 0.74, 0.78}
\definecolor{RYB2}{rgb}{0.18, 0.42, 0.41}
\definecolor{RYB6}{rgb}{0.85, 0.55, 0.13}
\definecolor{RYB5}{rgb}{0.69, 0.67, 0.66}
\definecolor{RYB3}{rgb}{0.72, 0.26, 0.06}
\definecolor{RYB4}{RGB}{251,220,127}

\pgfplotscreateplotcyclelist{newcolors}{
{RYB1,every mark/.append style={fill=RYB1,mark size={1}},mark=*},
{RYB3,every mark/.append style={fill=RYB3,mark size={1}},mark=*},
{RYB2,every mark/.append style={fill=RYB2,mark size={1}},mark=*},
{RYB5,every mark/.append style={fill=RYB5,mark size={1}},mark=*},
{RYB4,every mark/.append style={fill=RYB4,mark size={3}},mark=pentagon*},
{RYB6,every mark/.append style={fill=RYB6,mark size={4}},mark=10-pointed star},
{RYB7,every mark/.append style={fill=RYB7},mark=*},
}

\tikzset{font=\small}
\pgfplotsset{compat=1.18,every axis/.append style={
    label style={font=\small},
    tick label style={font=\small},
    width=7cm,
    height=5cm,
    cycle list name=newcolors,
    },
}

\newcommand\blfootnote[1]{%
  \begingroup
  \renewcommand\thefootnote{}\footnote{#1}%
  \addtocounter{footnote}{-1}%
  \endgroup
}
\begin{document}

\hypersetup{
  linkcolor=darkrust,
  citecolor=seagreen,
  urlcolor=darkrust,
  pdfauthor=author,
}

\begin{frontmatter}

\title{{\large\bfseries RoseNNa: A performant, portable library for neural network inference with application to computational fluid dynamics}}

\author{\vspace{-3ex}Ajay Bati}
\author{Spencer H.\ Bryngelson}
\ead{shb@gatech.edu}

\address{School of Computational Science \& Engineering, Georgia Institute of Technology, Atlanta, GA 30332, USA}

\date{}

\begin{abstract}
The rise of neural network-based machine learning ushered in high-level libraries, including TensorFlow and PyTorch, to support their functionality. Computational fluid dynamics (CFD) researchers have benefited from this trend and produced powerful neural networks that promise shorter simulation times. For example, multilayer perceptrons (MLPs) and Long Short Term Memory (LSTM) recurrent-based (RNN) architectures can represent sub-grid physical effects, like turbulence. Implementing neural networks in CFD solvers is challenging because the programming languages used for machine learning and CFD are mostly non-overlapping,  We present the roseNNa library, which bridges the gap between neural network inference and CFD. RoseNNa is a non-invasive, lightweight (1000 lines), and performant tool for neural network inference, with focus on the smaller networks used to augment PDE solvers, like those of CFD, which are typically written in C/C++ or Fortran. RoseNNa accomplishes this by automatically converting trained models from typical neural network training packages into a high-performance Fortran library with C and Fortran APIs. This reduces the effort needed to access trained neural networks and maintains performance in the PDE solvers that CFD researchers build and rely upon. Results show that RoseNNa reliably outperforms PyTorch (Python) and libtorch (C++) on MLPs and LSTM RNNs with less than 100 hidden layers and 100 neurons per layer, even after removing the overhead cost of API calls. Speedups range from a factor of about 10 and 2 faster than these established libraries for the smaller and larger ends of the neural network size ranges tested.
\end{abstract}

\end{frontmatter}

\blfootnote{Code available at: \url{https://github.com/comp-physics/roseNNa}}

\section{Introduction}\label{s:intro}

Deep learning has received considerable attention due to the availability of data and increasing computational power.
Computational fluid dynamics (CFD) practitioners have been developing neural-network-based models to enhance traditional closures models and numerical methods.
For example, \citet{fukami2020convolutional} implemented a convolutional autoencoder and multilayer perceptron (MLP) to speedup turbulence simulations, and \citet{zhu2021turbulence} showed how multiple artificial neural networks (ANNs) can model turbulence at high Reynolds numbers.
Of course, there are many other such examples.
These trained models show promising results but are often not integrated into high-performance solvers to deploy the model at scale.
Since the neural networks are typically constructed via Python-based learning libraries like PyTorch, it is unclear how to most efficiently introduce them into CFD and other PDE solvers written in low-level languages like C and Fortran. 

Researchers have proposed solutions to bridge the gap between the Python and HPC domains for deep learning.
Currently, the most frequently updated Fortran framework for this task is neural-Fortran~\cite{curcic2019parallel}, which supports building, training, and model parallelism in Fortran.
However, porting pre-trained neural networks to Fortran using this framework would require understanding their deep learning documentation, manually rewriting the model's architecture, and transferring the trained model's parameters.
Other attempts to solve the lack of deep learning support in HPC codebases also focus on manually specifying the architecture or converting neural network models from a single Python library to an HPC-amenable language.
For example, Fortran--Keras Bridge (FKB)~\cite{ott2020fortran} is derived from neural-Fortran and specializes in Keras-based models, NEURBT~\cite{bernal2015neurbt} focuses on neural networks for classification, FANN~\cite{nissen2003implementation} describes a C library for multilayer feed-forward networks, and SAGRAD~\cite{bernal2015sagrad} (like NEURBT) implements training in Fortran77. 

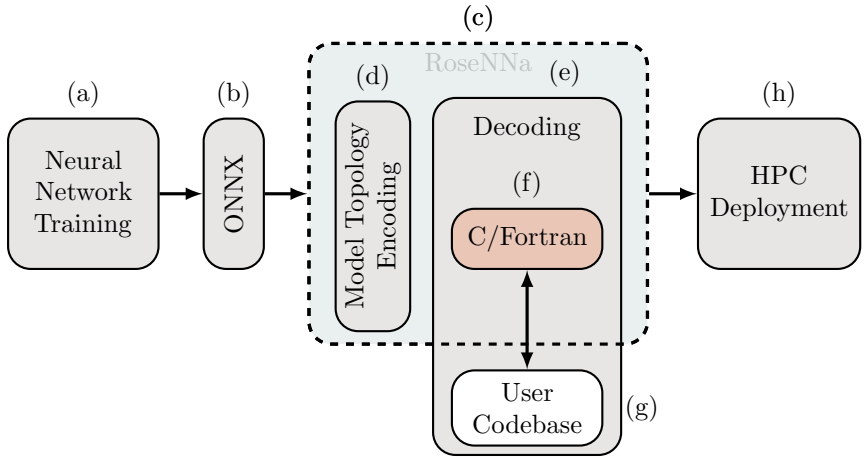
\begin{figure}[H]
    \centering
    \definecolor{lightblue}{rgb}{0.63, 0.74, 0.78}
\definecolor{seagreen}{rgb}{0.18, 0.42, 0.41}
\definecolor{orange}{rgb}{0.85, 0.55, 0.13}
\definecolor{silver}{rgb}{0.69, 0.67, 0.66}
\definecolor{rust}{rgb}{0.72, 0.26, 0.06}

\colorlet{lightsilver}{silver!30!white}
\colorlet{darkorange}{orange!75!black}
\colorlet{darksilver}{silver!65!black}
\colorlet{darklightblue}{lightblue!65!black}
\colorlet{darkrust}{rust!85!black}

\tikzstyle{trainingstyle} = [rectangle, 
minimum width=2cm, 
minimum height=2cm,
text centered,
text width=1.5cm,
draw=black,
thick,
rounded corners=0.3cm, 
fill=lightsilver, label=(a)]

\tikzstyle{onnxstyle} = [rectangle, 
minimum width=2cm, 
minimum height=0.8cm, 
text centered,
draw=black, 
thick,
rotate=90,
rounded corners=0.3cm,
fill=lightsilver, label=right:(b)]

\tikzstyle{mtestyle} = [rectangle, 
minimum width=2.7cm, 
minimum height=0.8cm, 
text centered, 
text width=2.8cm,
thick,
draw=black, 
rotate=90,
thick,
rounded corners=0.3cm,
fill=lightsilver, label=right:(d)]

\tikzstyle{decodingstyle} = [rectangle, 
minimum width=2cm, 
minimum height=0.8cm, 
text centered,
draw=black,
rotate=0,
thick,
rounded corners=0.3cm,
fill=rust!30, label=(f)]

\tikzstyle{eucstyle} = [rectangle, 
minimum width=2cm, 
minimum height=1cm,
text centered,
text width=1.55cm,
draw=black,
thick,
rounded corners=0.3cm, 
fill=white, label={[xshift=0.15cm]right:(g)}]

\tikzstyle{border1style} = [rectangle, 
minimum width=2.5cm, 
minimum height=4.75cm,
text centered,
text depth=4cm,
draw=black,
rounded corners=0.3cm, 
thick,
fill=lightsilver,
fill opacity=1, label={[xshift=0.5cm]above:(e)}]

\tikzstyle{border2style} = [rectangle, 
minimum width=4.5cm, 
minimum height=4cm,
text centered,
text depth=3.5cm,
draw=black,
dashed,
very thick,
rounded corners=0.3cm, 
fill=seagreen,
fill opacity=0.1,
text opacity=1,
label=(c)]

\tikzstyle{hpcstyle} = [rectangle, 
minimum width=2cm, 
minimum height=2cm,
text centered,
text width=1.9cm,
draw=black,
thick,
rounded corners=0.3cm, 
fill=lightsilver, label=(h)]

\tikzstyle{arrow} = [very thick,->,>=latex]
\tikzstyle{darrow} = [very thick,<->,>=latex]
    
\begin{tikzpicture}[node distance=1cm]
\node (training) [trainingstyle] {Neural Network Training};

\node (onnx) [onnxstyle, below of=training, yshift=-1cm] {ONNX};

\node (border2) [border2style, right of=onnx, xshift=2.25cm,yshift=0cm] {RoseNNa};

\node (mte) [mtestyle, below of=border2, yshift=2.4cm, xshift=-0.3cm] {Model Topology \\ Encoding};

\node (border1) [border1style, right of=mte, xshift=1.05cm, yshift=-0.8cm] {Decoding};

\node (decoding) [decodingstyle, below of=border1,xshift=0cm,yshift=1.5cm] {C/Fortran};

\node (euc) [eucstyle, below of=decoding, yshift=-1.25cm] {User \\ Codebase};

\node (hpc) [hpcstyle, right of=border2, xshift=3cm, yshift=0cm] {HPC \\ Deployment};

\node (border2) [border2style, right of=onnx, xshift=2.25cm,yshift=0cm, fill=none] {RoseNNa};

\draw [arrow] (training) -- (onnx);
\draw [arrow] (onnx) -- (border2);
\draw [darrow] (decoding) -- (euc);
\draw [arrow] (border2) -- (hpc);
\end{tikzpicture}
    \caption{
        RoseNNa (c) is a neural network converter that integrates into the inference process.
        It encodes the ONNX-converted neural network and transforms it into performant Fortran code with C and Fortran APIs.
        The user provides the components outside of (c).
    }
    \label{fig:pipeline}
\end{figure}

We avoid these drawbacks via a fast and non-intrusive automatic conversion tool.
This manuscript presents an open-source library called RoseNNa that achieves these tasks.
RoseNNa is available under the MIT license at \url{github.com/comp-physics/roseNNa}. 
As shown in \cref{fig:pipeline}, RoseNNa encodes pre-trained neural networks from common machine learning libraries via an ONNX-backend and uses fypp, a Python-to-Fortran metaprogramming language, to generate the library.

RoseNNa targets inference of the artificial neural networks used for PDE- and CFD-based modeling, supporting commonly used architectures and activation functions in these areas.
These network architectures were revealed via a literature survey of about 25 papers that used neural networks for modeling and numerics tasks.
This survey found that MLP implementations consisted of at most 6 hidden layers, and 85\% of them have fewer than 15~hidden layers (or dimensionality), and the remaining fraction having fewer than 100~neurons per layer\footnote{Search terms: ``mlp turbulence modeling,'' ``mlp cfd solver,'' ``multilayer perceptron computational fluid dynamics''}.
We also reviewed 10 articles using LSTM architectures for similar purposes.\footnote{Search terms: ``subgrid closures rnn,'' ``lstm rnn cfd solver,'' ``lstm turbulence closure,'' ``lstm rnn rans cfd''}
90\% of implementations use fewer than 64-time steps in the memory layer and a have a hidden dimension smaller than 32.
These numbers provide the architectures that RoseNNa should support with high performance.
The results are also consistent with expectations: PDE solvers on discretized grids with many elements that require many iterations (or time steps) cannot afford the evaluation of large neural networks since they involve relatively many floating point operations.

This manuscript discusses the methodology surrounding RoseNNa and example applications to CFD. 
\Cref{s:design} introduces the architecture of RoseNNa that enables its flexibility and speed.
\Cref{s:api} describes its user-friendly interface and design, and
\Cref{s:results} discusses RoseNNa's performance results against Python-based libraries for popular CFD architectures.
We conclude in \cref{s:conclusions} with a discussion of the primary results and use cases of the RoseNNa tool.

\section{Design strategy}\label{s:design}

\subsection{Design options}

RoseNNa follows two main processes: read and interpret the Python-native model (encode, \cref{fig:pipeline}~(d)) and reconstruct it in Fortran/C (decode, \cref{fig:pipeline}~(e)).
The tool decomposes key aspects of a neural network to define its structure: trained weights (values and dimensions), layer functionality, activation functions, and the order of layer connections. Using this encoded information, RoseNNa can reconstruct the functionality of a neural network in Fortran/C.

Users first convert their model to a unified format via ONNX~\citep{bai2019} (\cref{fig:pipeline}~(b)), a library that provides interoperability between machine learning libraries, including sklearn~\citep{pedregosa2011scikit}, PyTorch~\citep{paszke2019pytorch}, TensorFlow~\citep{abadi2016tensorflow}, and Caffe~\citep{jia2014caffe}.
These libraries share common characteristics in their intermediary representations and the functionality of a neural network model.
Still, they differ in their layer encoding. ONNX unifies these differences.
    
The ONNX-interpreted model is decoded using fypp (\cref{fig:pipeline}~(f)), a Python-based pre-processor for Fortran codes.
The activation functions (Tanh, ReLU, Sigmoid), model layers (LSTM, convolutions, pooling layers, MLP architectures, and more), and data structures holding model weights are first extracted via fypp and then stored in RoseNNa, specifically in Fortran code.
This ensures no speed is lost to reading in needed values while conducting inference.
ONNX encodes the model's structure while RoseNNa restores its graph interpretation using fypp.

Alternative solutions to Python-to-HPC model conversion are also viable.
For example, Python functionality can be integrated into Fortran by running an instance of Python or exposing a model's outputs through APIs.
These attempts, however, are susceptible to cascading overhead time issues.
The library we present, RoseNNa, removes this overhead and enables quick HPC deployment, features that CFD practitioners require for running simulations.
The power of RoseNNa comes from its internal management of neural networks, ONNX backend, and Fortran/C support.

Like established linear algebra libraries like BLAS and LAPACK, RoseNNa is a Fortran library.
This is an appropriate fit since the library's focus is fast evaluation of rather simple mathematical functions, like small matrix--matrix and matrix--vector products.
In addition, with recent updates to the language, Fortran can be readily linked to C, which is also often used for these applications.
Fortran compilers are well optimized and can efficiently handle small matrix--matrix multiplies.
We found that the optimized Python-based inference speeds for smaller model architectures are similar to the speeds seen in RoseNNa, as shown in \cref{fig:times-mlp} and \cref{fig:times-lstm}.

\subsection{ONNX}

Open Neural Network Exchange (ONNX)~\citep{bai2019} is an open-source artificial intelligence (AI/ML) ecosystem that allows for interoperability between preexisting machine learning libraries and provides inference optimizations.
During the pre-processing stage, RoseNNa encodes the neural network model.
This entails parsing and storing each layer's order, weights, dimensions, and other functionality in the library.
We use ONNX to unify differences between neural network model interpretations and establish a common parser that can be optimized at compile time.
Users can convert their model to the ONNX due to its widespread interoperability support.
ONNX is often used in research and industry.
For example, \citet{someki2022espnet} used ONNX to unify functionality support and \citet{moreno2020jedi} converted a PyTorch model to a TensorFlow graph for compatibility with testing software.
Like \citet{rodriguez2020deep}, RoseNNa reconstructs models from deep-learning libraries, enabling model designers to keep their native framework.

\subsection{Metaprogramming}

The transition from model topology encoding \cref{fig:pipeline}~(d) to the decoding stage in \cref{fig:pipeline}~(e) is performed by a Python-based Fortran pre-processor called fypp~\citep{balint_aradi_2020_3605649}.
In our implementation, fypp translates a neural network's properties into Fortran code \textit{before} compile-time, thus exposing compiler optimizations.
This decoding process is unique to each neural network, and so is re-run for different neural network models.

After interpreting the Python-native model, we store its features and important variable definitions in fypp files.
This encoding process stores the layers, activation functions, and weight parameters while preserving their order.
We record these layers' specific options, including whether transposing is required and hyperparameter constants.
RoseNNa tracks changes in matrix shapes, allowing it to define variables with their appropriate dimensions in Fortran explicitly.
It also stores the output names of each layer so they can be referenced during the decoding phase in Fortran.
To increase readability, these output names are only defined when the input undergoes dimension changes.
The decoding stage (\cref{fig:pipeline}~(f)) references each component described above.
Using fypp, the layers are defined in order with their respective weights, constants, and other supplementary options.

\begin{figure}
    \centering
    \begin{minipage}[t]{0.43\textwidth}
    \begin{lstlisting}[caption={Fypp code to generate a linear layer}]
#: if tup[0] == 'Gemm'
!===Gemm Layer===
call linear_layer(${tup[1][0]}$,         & 
  linLayers(${layer_dict[tup[0]]}$), &
  ${1-tup[1][1]}$)
    \end{lstlisting}
    \end{minipage}%
    \qquad 
    \begin{minipage}[t]{0.09\textwidth}
    \centering
    \vspace{1.25cm}
    \Large$\xrightarrow{\normalsize\,\,\mathrm{fypp}\,\,}$
    \end{minipage}
    \qquad
    \begin{minipage}[t]{0.33\textwidth}
    \begin{lstlisting}[language=Fortran, caption=Corresponding Fortran]
!===Gemm Layer===
call linear_layer(input, & 
  linLayers(1),0)
    \end{lstlisting}
    \end{minipage}
\end{figure}

\subsection{RoseNNa capabilities}

RoseNNa was designed to support a broad range of neural network architectures in CFD. 
As discussed in \cref{s:intro}, these primarily include MLPs and LSTM RNNs.
RoseNNa also supports other architectures, such as convolutional and pooling layers, which are generally popular and could become more broadly used in CFD solvers in the future.
One can expand RoseNNa for different architectures and activation functions as needed.
Adding these new features to the tool requires only a basic understanding of the architecture functionality, how ONNX encodes it, and following the RoseNNa contributor's guide for implementation.

\section{User interface}\label{s:api}

The user will have access to all files that make up the library.
RoseNNa is designed for straightforward and non-intrusive integration in existing codebases.
As described in the pipeline of \cref{fig:pipeline}~(b), the only required input to RoseNNa is an ONNX-format pre-trained neural network model.
Simple pre-processing using the metaprogramming language fypp reconstructs the neural network, creating a custom Fortran file with an organized subroutine defining the model's structure.
Compiling all core files and the fypp-transcribed file creates a library that can be linked with an existing code (\cref{fig:pipeline}~(g)).

Using RoseNNa in C, except for defining headers for certain function calls, follows the same procedure.
To use this library, one imports RoseNNa, which automatically reads and initializes the parameters encoded from the trained neural network, and then calls the model's forward subroutine with the same inputs as the native model.
\Cref{lst:RoseNNa} shows this lightweight approach.

\begin{lstlisting}[language=Fortran, caption=Example Fortran90+ program invoking RoseNNa.,label={lst:RoseNNa},float]
program example
  use rosenna !import the library
  implicit none
  real(c_double), dimension(1,2) :: inputs
  real(c_double), dimension(1,3) :: output

  inputs = reshape((/1.0, 1.0/), (/1, 2/), order=[2, 1])
  call initialize() !initialize/load in the weights
  call use_model(inputs, output) !conduct inference, store output
end program
\end{lstlisting}

\section{Results}\label{s:results}

\subsection{Flexibility and portability}

With only a few library calls, RoseNNa can be readily integrated into existing programs.
It can interface with commonly used machine learning libraries and be linked to Fortran and C, the most popular languages in CFD. 
RoseNNa can dynamically reconstruct neural networks and avoids any manual intervention.
RoseNNa can also represent attributes of deep learning models: Layers, activation functions, important constants, and more.
RoseNNa caters to smaller neural networks (MLPs and LSTMs) and supports around $90\%$ of the most popular architectures and activation functions used in CFD research.
Its simple user interface and ability to interpret ONNX-format models enable the conversion of a massive pool of promising neural networks in CFD.

\subsection{Performance on example cases}

We ran tests for CFD's most commonly used architectures, LSTMs and MLPs, to compare inference performance differences between RoseNNa and PyTorch. 
We compare RoseNNa's performance to that of PyTorch because it is a representative and widely used deep learning library. 

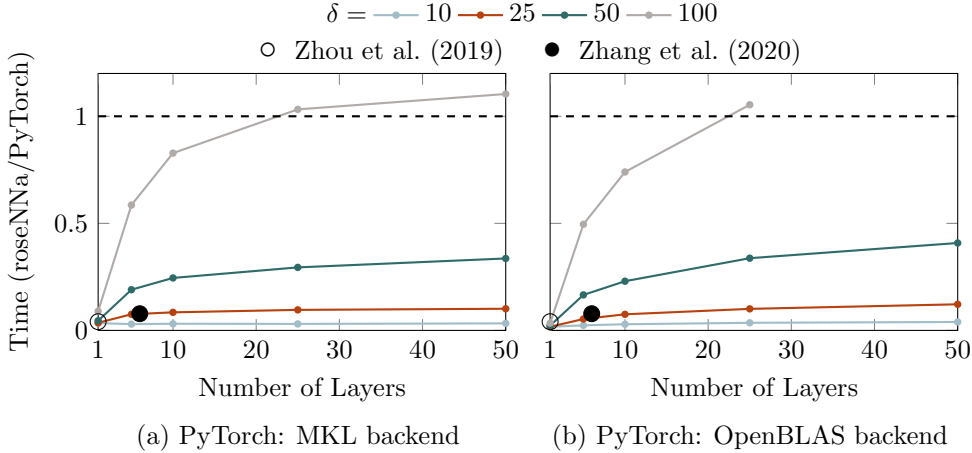
\begin{figure}[t]
    \centering
    \begin{subfigure}[t]{0.45\linewidth}
        \centering
        \begin{tikzpicture}
\begin{axis}[
    xlabel={Number of Layers},
    ylabel={Time (roseNNa/PyTorch)},
    ymax=1.2,
    xmin=1,
    ymin=0,
    xmax=50,
    legend style={
        at={(1.04,1.15)},
        inner sep=1pt,
        anchor=south,
        legend columns=5,
        legend cell align={left},
        draw=none,
        fill=none,
    },
    extra x ticks={1},
    clip=false,
]
\node [] at (70,1.3) {{\Large\textbullet}\,\, Zhang et al.\ (2020)};
\node [] at (35,1.3) {{\Large\textopenbullet}\,\, Zhou et al.\ (2019)};
\node [] at (25,-0.5) {(a) PyTorch: MKL backend};
\addlegendimage{empty legend};
\addlegendentry{$\delta=$};
\addlegendentry{10}
\addplot+[thick,
    ]
    coordinates {
    (1.0, 3.394055e-02)
    (5.0, 2.937064e-02)
    (10.0, 3.102483e-02)
    (25.0, 3.025623e-02)
    (50.0, 3.287788e-02)
    };
    
\addlegendentry{25}
\addplot+[thick,
    ]
    coordinates {
    (1.0, 3.558317e-02)
    (5.0, 7.524440e-02)
    (10.0, 8.443477e-02)
    (25.0, 9.588120e-02)
    (50.0, 1.011111e-01)
    };
    
\addlegendentry{50}
\addplot+[thick,
    ]
    coordinates {
    (1.0, 4.515761e-02)
    (5.0, 1.900203e-01)
    (10.0, 2.450138e-01)
    (25.0, 2.942578e-01)
    (50.0, 3.359562e-01)
    };

\addlegendentry{100}
\addplot+[thick,
    ]
    coordinates {
    (1.0, 8.960639e-02)
    (5.0, 5.853618e-01)
    (10.0, 8.280444e-01)
    (25.0, 1.032249e+00)
    (50.0, 1.104072e+00)
    };
\addplot[mark=o,mark size=3pt,
    color=black,pin={[red]60:Hey!}
    ]
    coordinates {
    (1, 0.040669748856)
    };

\addplot[mark=*,mark size=3pt,
    ]
    coordinates {
    (6,0.07793309457)
    };
\addplot[thick,dashed,black,domain=1:50] plot {1};    

\end{axis}
\end{tikzpicture}
    \end{subfigure}
    \hspace{-0.5in}
    \begin{subfigure}[t]{0.45\linewidth}
        \centering
        \begin{tikzpicture}
\begin{axis}[
    xlabel={Number of Layers},
    ymax=1.2,
    ymin=0,
    xmax=50,
    xmin=1,
    yticklabels={,,},
    extra x ticks={1},
    clip=false,
]
\node [] at (25,-0.5) {(b) PyTorch: OpenBLAS backend};
\addplot+[thick,
    ]
    coordinates {
    (1.0, 1.678150e-02)
    (5.0, 2.362474e-02)
    (10.0, 2.873442e-02)
    (25.0, 3.553864e-02)
    (50.0, 3.934824e-02)
    };
    
\addplot+[thick]
    coordinates {
    (1.0, 1.688242e-02)
    (5.0, 5.318252e-02)
    (10.0, 7.514234e-02)
    (25.0, 1.006175e-01)
    (50.0, 1.218839e-01)
    };
    
\addplot+[thick,
    ]
    coordinates {
    (1.0, 2.245593e-02)
    (5.0, 1.657628e-01)
    (10.0, 2.297416e-01)
    (25.0, 3.372285e-01)
    (50.0, 4.081896e-01)
    };

\addplot+[thick]
    coordinates {
    (1.0, 3.376717e-02)
    (5.0, 4.955171e-01)
    (10.0, 7.399281e-01)
    (25.0, 1.053850e+00)
    };
\addplot[mark=o,mark size=3pt,
    color=black,pin={[red]60:Hey!}
    ]
    coordinates {
    (1, 0.040669748856)
    };

\addplot[mark=*,mark size=3pt,
    color=black,
    ]
    coordinates {
    (6,0.07793309457)
    };
\addplot[thick,dashed,black,domain=1:50] plot {1};    

\end{axis}
\end{tikzpicture}
     \end{subfigure}
     \caption{
        Multilayer perceptron (MLP) time comparison (RoseNNa versus PyTorch).
        $\delta$ represents a specific hidden size (neurons per layer), and the x-axis represents the depth (number of hidden layers).
        Random activation functions (ReLu, Tanh, Sigmoid) were chosen for each MLP and assigned to each hidden layer.
        }
     \label{fig:times-mlp}
\end{figure}

\begin{figure}[t]
    \centering
    \begin{subfigure}[t]{0.45\linewidth}
        \centering
        \begin{tikzpicture}
\begin{axis}[
    xlabel={Sequence Length},
    ylabel={Times (roseNNa/PyTorch)},
    ymax=0.7,
    ymin=0,
    xmin=1,
    xmax=100,
    legend style={
        at={(1.04,1.15)},
        inner sep=1pt,
        anchor=south,
        legend columns=5,
        legend cell align={left},
        draw=none,
        fill=none,
    },
    extra x ticks={1},
    clip=false,
]
\node [] at (75,0.75) {{\Large\textopenbullet}\,\, Srinivasan et al.\ (2019)};
\node [] at (145,0.75) {{\Large\textbullet}\,\, Li et al.\ (2020)};
\node [] at (50,-0.28) {(a) PyTorch: MKL backend};
\addlegendimage{empty legend};
\addlegendentry{$\lambda=$};
\addlegendentry{3}
\addplot+[thick,
    ]
    coordinates {
    (2.0, 3.369956e-02)
    (5.0, 3.375745e-02)
    (10.0, 3.431406e-02)
    (50.0, 3.380869e-02)
    (100.0, 3.331934e-02)
    };
    
\addlegendentry{25}
\addplot+[thick,
    ]
    coordinates {
    (2.0, 9.419549e-02)
    (5.0, 1.194920e-01)
    (10.0, 1.358244e-01)
    (50.0, 1.602004e-01)
    (100.0, 1.647159e-01)
    };
    
\addlegendentry{50}
\addplot+[thick,
    ]
    coordinates {
    (2.0, 1.680867e-01)
    (5.0, 2.294296e-01)
    (10.0, 2.824407e-01)
    (50.0, 3.318823e-01)
    (100.0, 3.456150e-01)
    };

\addlegendentry{75}
\addplot+[thick,
    ]
    coordinates {
    (2.0, 2.568633e-01)
    (5.0, 3.538743e-01)
    (10.0, 4.204438e-01)
    (50.0, 5.212900e-01)
    (100.0, 5.363372e-01)
    };
\addplot[mark=o,mark size=3pt,
    ]
    coordinates {
    (5,0.01037739356)
    };

\addplot[mark=*,mark size=3pt,
    ]
    coordinates {
    (25, 0.129327662217)
    };

\end{axis}
\end{tikzpicture}
    \end{subfigure}
    \hspace{-0.5in}
    \begin{subfigure}[t]{0.45\linewidth}
        \centering
        \begin{tikzpicture}
\begin{axis}[
    xlabel={Sequence Length},
    ymax=0.7,
    xmax=100,
    xmin=1,
    ymin=0,
    yticklabels={,,},
    extra x ticks={1},
    clip=false,
]
\node [] at (50,-0.28) {(b) PyTorch: OpenBLAS backend};
\addplot+[thick,
    ]
    coordinates {
    (2.0, 2.069871e-02)
    (5.0, 2.505159e-02)
    (10.0, 2.953171e-02)
    (50.0, 3.614491e-02)
    (100.0, 3.822778e-02)
    };
\addplot+[thick,
    ]
    coordinates {
    (2.0, 5.862273e-02)
    (5.0, 9.027452e-02)
    (10.0, 1.214624e-01)
    (50.0, 1.707765e-01)
    (100.0, 1.804598e-01)
    };
\addplot+[thick,
    ]
    coordinates {
    (2.0, 1.109874e-01)
    (5.0, 1.855829e-01)
    (10.0, 2.522165e-01)
    (50.0, 3.763113e-01)
    (100.0, 4.022749e-01)
    };
\addplot+[thick,
    ]
    coordinates {
    (2.0, 1.679841e-01)
    (5.0, 2.770341e-01)
    (10.0, 3.796703e-01)
    (50.0, 5.753655e-01)
    (100.0, 6.078791e-01)
    };
\addplot[mark=o,mark size=3pt,
    ]
    coordinates {
    (5,0.01037739356)
    };

\addplot[mark=*,mark size=3pt,
    ]
    coordinates {
    (25, 0.129327662217)
    };
\end{axis}
\end{tikzpicture}
    \end{subfigure}
    \caption{
        Long Short-Term Memory (LSTM) time comparison (RoseNNa/PyTorch).
        The horizontal axis is the number of time steps (depth), and $\lambda$ is the hidden dimension size.
        All the typical operations and activation functions were incorporated into the timing of the LSTM cells.
    }
    \label{fig:times-lstm}
\end{figure}
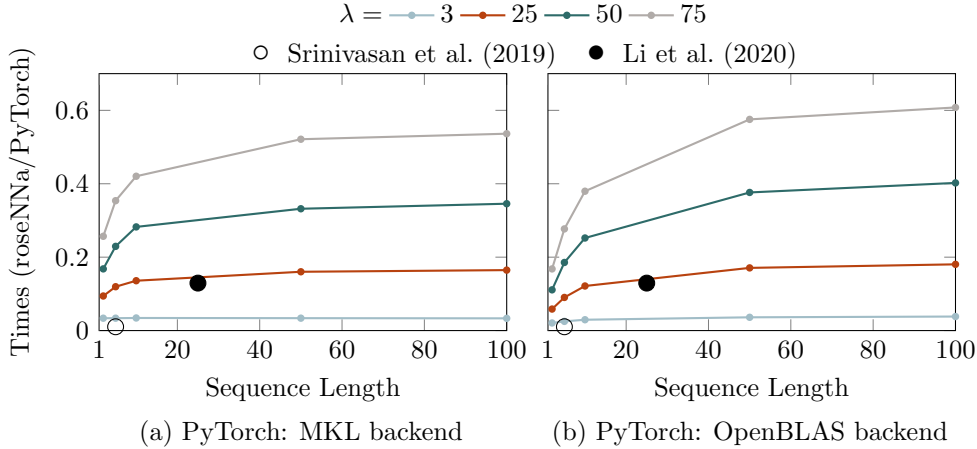

We used a single core of an Intel Xeon Gold 6226 CPU to run the following tests. 
We ran 100 tests in PyTorch on a single thread for each data point using randomly initialized weights. 
The same models curated in PyTorch were converted to and tested in RoseNNa. 
Then, we took the ratio of the medians of the 100 RoseNNa and 100 PyTorch times. 
This process was repeated 25 times for each point in \cref{fig:times-mlp,fig:times-lstm,fig:times-mlp-cpp}.

Results for MLPs are important due to their widespread usage in CFD solvers.
Their straightforward architectures and computation also allow for unproblematic conversions.
Most MLPs used in CFD are shallow to enable reasonable computation runtimes.
Based on this and the results of our literature survey, MLPs used to solve large PDE systems like those of CFD fall within the axis limits of \cref{fig:times-mlp}.

\Cref{fig:times-mlp} shows that tests fall under a RoseNNa-to-PyTorch time ratio of one, indicating RoseNNa's quicker inference speeds.
The ratio stays near one even for large examples such as 50 neurons and a depth of 100, which is uncommon for CFD applications.
We further tested RoseNNa's inference speeds against a different PyTorch backend for consistency and to ensure we compared against the fastest version of PyTorch. 
Therefore, \cref{fig:times-mlp}~(b) represents the same tests run on PyTorch with an OpenBLAS backend instead of MKL. 
RoseNNa is $10\%$ faster (averaged over all 25 test cases) using this backend, but the results still fall below a one-time ratio for most CFD use cases. 
However, with OpenBLAS, larger architectures entail increasingly slower times.

Small-scale LSTM--RNN architectures are also often used in CFD applications.
\Cref{fig:times-lstm} shows tests conducted at different depths and hidden dimension sizes to demonstrate where RoseNNa falls compared to PyTorch inference speeds.
Most CFD-based LSTMs' architectures are located below the one RoseNNa-to-PyTorch time ratio.
Compared with an OpenBLAS implementation of PyTorch, RoseNNa seems to be $10\%$ faster on average.
Larger architectures lead to a slower inference time ratio as expected.

\Cref{fig:times-mlp} and \cref{fig:times-lstm} incorporate published examples of LSTMs and MLPs.
All four test cases lie in the bottom left corner of the graph since they are shallow architectures.
A simple conversion from PyTorch or TensorFlow to ONNX allowed us to pass the model through RoseNNa's pipeline.
Despite their shallow architectures, these papers reported promising results across CFD modeling tasks broadly.
For MLPs, \citet{zhang2020large} proposed combining an artificial neural network with a flamelet-generated manifold to solve a memory issue.
\citet{zhou2019subgrid} developed a new SGS model for large-eddy simulation (LES), showing significant improvements over the conventional models.
For LSTMs, \citet{srinivasan2019predictions} found this architecture outperformed MLPs in predicting turbulent statistics in temporally evolving turbulent flows.
Lastly, \citet{li2020nonlinear} uses LSTMs to develop a reduced-order modeling of a wind-bridge interaction system.

\subsection{Comparison to a lower-level implementation}

\begin{figure}
    \centering
    \begin{tikzpicture}
\begin{axis}[
    xlabel={Number of Layers},
    ylabel={Time (roseNNa/LibTorch)},
    ymax=2,
    xmin=1,
    ymin=0,
    xmax=50,
    legend style={
        at={(0.5,1.05)},
        inner sep=1pt,
        anchor=south,
        legend columns=5,
        legend cell align={left},
        draw=none,
        fill=none,
    },
    extra x ticks={1},
    clip=false,
]

\addlegendimage{empty legend};
\addlegendentry{$\delta=$};
\addlegendentry{10}
\addplot+[thick,
    ]
    coordinates {
    (1.0, 1.310683E-01)
    (5.0,     7.361645E-02)
    (10.0,     6.515091E-02)
    (25.0,     5.522680E-02)
    (50.0,     5.677159E-02)
    };
    
\addlegendentry{25}
\addplot+[thick,
    ]
    coordinates {
    (1.0, 1.732140E-01)
    (5.0, 1.772946E-01)
    (10.0, 1.692639E-01)
    (25.0, 1.733225E-01)
    (50.0, 1.764597E-01)
    };
    
\addlegendentry{50}
\addplot+[thick,
    ]
    coordinates {
    (1.0, 1.735221E-01)
    (5.0, 4.389012E-01)
    (10.0, 5.194365E-01)
    (25.0, 5.651535E-01)
    (50.0, 5.781099E-01)
    };

\addlegendentry{100}
\addplot+[thick,
    ]
    coordinates {
    (1.0, 2.183232E-01)
    (5.0, 1.470857E+00)
    (10.0, 1.709954E+00)
    (25.0, 1.831430E+00)
    (50.0, 1.830916E+00)
    };

\addplot[thick,dashed,black,domain=1:50] plot {1};    

\end{axis}
\end{tikzpicture}
    \caption{
        Multilayer perceptron (MLP) model time comparison (RoseNNa/libtorch). $\delta$ is the hidden size, and the horizontal axis is the number of layers. 
        Libtorch is PyTorch's C++ API. 
        The same scheme for testing the RoseNNa to PyTorch speed ratio for MLPs was used for these tests.
    }
    \label{fig:times-mlp-cpp}
\end{figure}
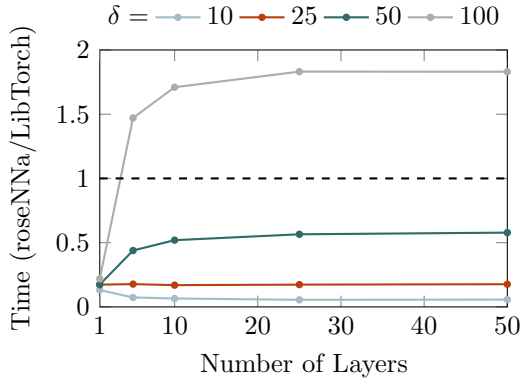

Another approach to reducing Python overhead is to use a library's C/C++ API if supported. 
For example, PyTorch has a (beta) fully-native C++ API called libtorch that provides access to most PyTorch functionality~\citep{paszke2019pytorch}. 
\Cref{fig:times-mlp-cpp} represents comparison tests run on the same architectures as \cref{fig:times-mlp} but against libtorch. 
Most architecture sizes are inferred faster via RoseNNa, in particular the smaller ones relevant to CFD simulation.
Larger neural networks, most of which are outside the CFD scope, are still slower but near RoseNNa's speed, even for sizes as large as 15 layers of 100 neurons. 
Libtorch makes up some of the RoseNNa--PyTorch speed difference for larger cases, but there are still potential issues with relying upon the Torch C++ API (and other exposed backend APIs).
For example, libtorch support is liable to change, which is stated directly on the Torch website.
It also only provides a C++ API, thus requiring more work, like a shim layer, for use in Fortran codebases than RoseNNa.

Python-based overhead might explain part of the time discrepancy of \cref{fig:times-mlp} and \cref{fig:times-lstm}, but the main contribution towards the speedup is RoseNNa's compile-time optimization and Fortran implementation.
These results show RoseNNa's computationally viability for ML-enhanced CFD.
For the larger architectures in \cref{fig:times-mlp,fig:times-lstm,fig:times-mlp-cpp} that were slower in RoseNNa, one can implement large matrix--matrix multiplies and other expensive calls via optimized linear algebra libraries like BLAS/LAPACK. 
However, based on our literature survey, these larger neural networks fall outside the CFD (and PDE-solver) scope RoseNNa focuses on.
With no external dependencies, the RoseNNa library is lightweight and can be readily incorporated into existing PDE solvers.

\section{Conclusions}\label{s:conclusions}

This paper describes the design, application, and viability of RoseNNa, a neural network conversion tool for CFD codebases.
It can encode a neural network's features using ONNX, a Python-based library we use to unify machine learning libraries.
With a Python-powered pre-processor, fypp, RoseNNa decodes the model in Fortran.
We present this tool as an alternative to manually defining neural networks in Fortran or re-implementing existing libraries' ML features.
In three speed comparison benchmarks we conducted (RoseNNa/MKL, RoseNNa/OpenBLAS, RoseNNa/libtorch), RoseNNa's application in the CFD domain seemed to be promising and a more reliable alternative to low-level implementations of PyTorch, TensorFlow, or other Python-based machine learning libraries.
RoseNNa supports many popular features and establishes a streamlined process for increasing its breadth.

RoseNNa presents useful benefits for neural network conversion and inference.
First, it supports the conversion from Python machine-learning libraries via ONNX.
It is also simple to use.
As shown in \cref{lst:RoseNNa}, a few API calls enable inference.
Lastly, RoseNNa is a lightweight tool, enabling integration and minimal intrusiveness in existing and (potentially large) CFD codebases.
The library is compiled for the neural network and linked to existing code.

RoseNNa's future lies in improving performance, adding functionality to existing architectures, and expanding to new, popular features. 
With the pipeline of \cref{fig:pipeline}, contributors can incorporate any needed feature. 
We have created an in-depth manuscript about our current methodology and how new contributions can be feasibly integrated (accessible at \url{github.com/comp-physics/roseNNa}). 
We provide steps describing which files to modify and examples, with documentation, of their functions and variables. 
Any new changes can be verified via the testing pipeline, allowing contributors to add new features efficiently.

\section*{Acknowledgements}

This work used Bridges2 at the Pittsburgh Supercomputing Center through allocation TG-PHY210084 (PI Spencer Bryngelson) from the Advanced Cyberinfrastructure Coordination Ecosystem: Services \& Support (ACCESS) program, which is supported by National Science Foundation grants \#2138259, \#2138286, \#2138307, \#2137603, and \#2138296.
SHB also acknowledges the resources of the Oak Ridge Leadership Computing Facility at the Oak Ridge National Laboratory, which is supported by the Office of Science of the U.S.\ Department of Energy under Contract No.\ DE-AC05-00OR22725. 
SHB acknowledges support from the Office of the Naval Research under grant N00014-22-1-2519 (PM Dr.\ Julie Young).
This research was supported in part through research cyberinfrastructure resources and services provided by the Partnership for an Advanced Computing Environment (PACE) at the Georgia Institute of Technology, Atlanta, Georgia, USA.

\bibliographystyle{model1-num-names}
\bibliography{main.bib}

\end{document}